\documentclass[journal=jctcce,manuscript=article]{achemso}

\usepackage{chemformula} 
\usepackage[T1]{fontenc} 

\usepackage{bm}          
\usepackage{amssymb,amsmath}
\usepackage{natbib}
\usepackage{ctable}
\usepackage{algorithm}
\usepackage{regexpatch}
\usepackage{algpseudocode}
\usepackage{multirow}
\usepackage{mathtools}
\usepackage{mathrsfs}
\usepackage{contour}
\usepackage{hyperref} 

\newcommand*{\eq}[1]{Eq.(\ref{#1})}
\newcommand*{\fig}[1]{Fig.\ref{#1}}
\newcommand*\Bell{\ensuremath{\boldsymbol\ell}} 

\author{Ahai Chen}
\affiliation{Universit{\'e} Paris-Saclay, UVSQ, CNRS, CEA, Maison de la Simulation, 91191, Gif-sur-Yvette, France}
\alsoaffiliation{Universit{\'e} Paris-Saclay, CNRS, Institut de Chimie Physique, UMR-CNRS 8000, 91405 Orsay, France}

\author{David M.\ Benoit }
\affiliation{E.A.\ Milne Centre for Astrophysics, The University of Hull, Cottingham Road, Kingston upon Hull HU6 7RX, UK}

\author{Yohann Scribano}
\affiliation{Laboratoire Univers et Particules de Montpellier, Universit\'e de Montpellier, UMR-CNRS 5299, 34095 Montpellier Cedex, France}

\author{Andr\'e Nauts}
\affiliation{Institute of Condensed Matter and Nanosciences (NAPS), Universit\'e Catholique de Louvain, Louvain-la-Neuve, Belgium}
\alsoaffiliation{Universit{\'e} Paris-Saclay, CNRS, Institut de Chimie Physique, UMR-CNRS 8000, 91405 Orsay, France}

\author{David Lauvergnat}
\email{david.lauvergnat@universite-paris-saclay.fr}
\affiliation{Universit{\'e} Paris-Saclay, CNRS, Institut de Chimie Physique, UMR-CNRS 8000, 91405 Orsay, France}

\title[]
  {A Smolyak algorithm adapted to a system-bath separation: application to an encapsulated molecule with large amplitude motions}

\begin{document}

\begin{tocentry}
\includegraphics[width=0.880\textwidth]{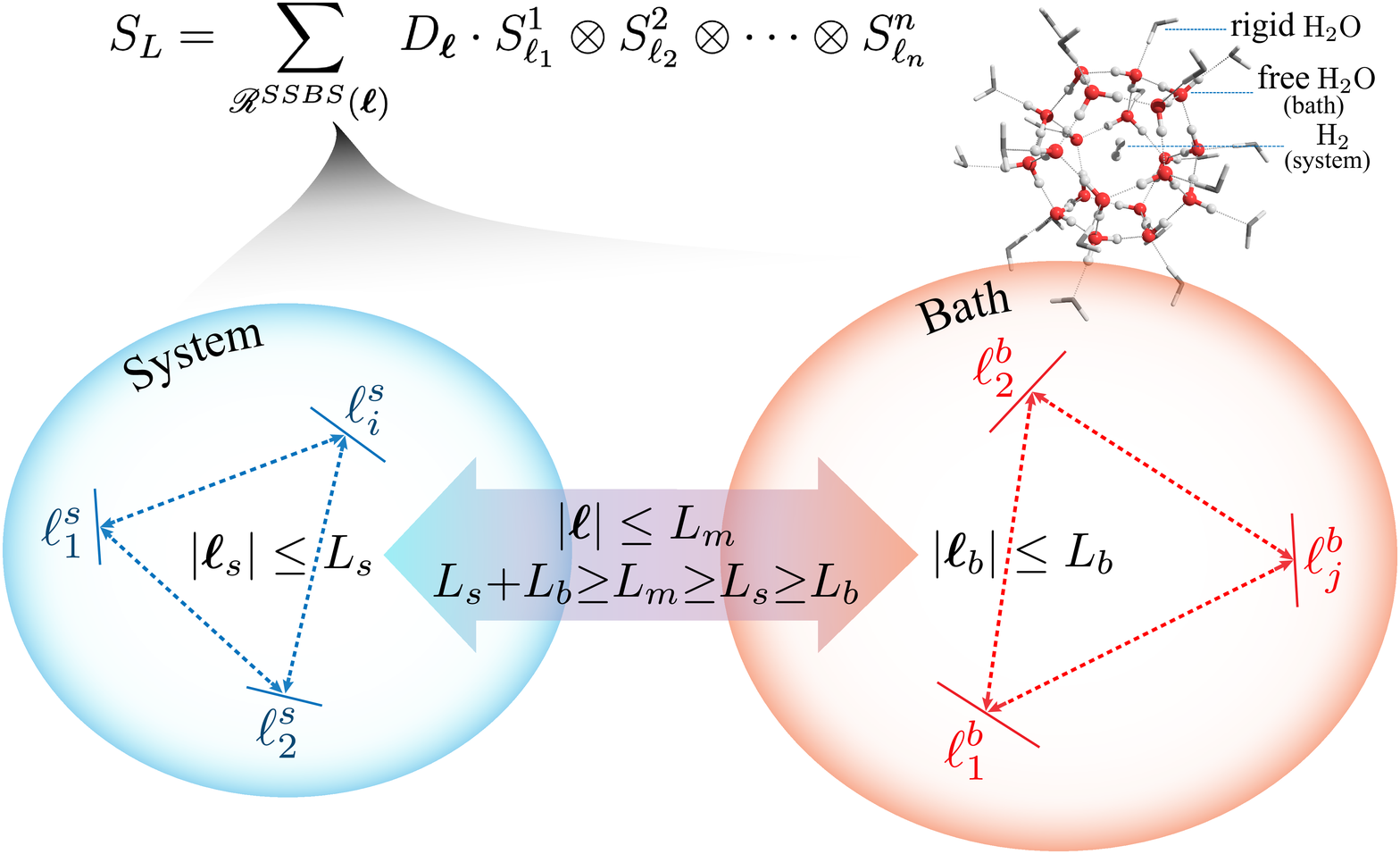} 




\end{tocentry}

\begin{abstract}
A Smolyak algorithm adapted to system-bath separation is proposed for rigorous quantum simulations.
This technique combines a sparse grid method with the system-bath concept in a specific configuration without limitations on the form of the Hamiltonian, thus achieving a highly efficient convergence of the excitation transitions for the ``system" part. Our approach provides a general way to overcome the perennial convergence problem for the standard Smolyak algorithm and enables the simulation of floppy molecules with more than a hundred degrees of freedom. The efficiency of the present method is illustrated on the simulation of H$_2$ caged in an sII clathrate hydrate including two kinds of cage modes. The transition energies are converged by increasing the number of normal modes of water molecules.
Our results confirm the triplet splittings of both translational and rotational ($j=1$) transitions of the H$_2$ molecule. Furthermore, they show a slight increase of the translational transitions with respect to the ones in a rigid cage.
\end{abstract}


\section{Introduction}

The study of quantum systems coupled to a complex host, for instance, the guest-host complexes~\cite{Cram1974_Host_Guest,Lehn1985_Host_Guest,Stucky1990_Host_Guest,Valdes2018_C60_H2O}, has attracted strong interest in recent years~\cite{review_Rotter2015_OQS,review_Alonso2017_OQS,review_Weimer2021_OQS}, spanning the fields of fundamental quantum physics, physical chemistry, quantum optics, etc. Some of these problems go beyond a structureless environment description, leading to the emergence of more accurate non-Markovian quantum models based on a system-bath approach~\cite{review_Alonso2017_OQS,review_Weimer2021_OQS,Soley2022_Tensor_Chebyshev}.
Given the well-known exponential scaling of the problem, the bath is usually described using effective approaches, e.g. a collection of harmonic oscillators, with a reconstructed effective system-bath coupling, e.g. the spectral density formulation~\cite{Alvermann2009_Smolyak_sys_bath,bonfanti2016_OQS,review_Alonso2017_OQS,Strathearn2018_OQS_NonM}. These approaches are limited, to some extent, to specifically tailored systems. Meanwhile, 
a variety of multi-configurational quantum dynamics method have been proposed, for example the well-known multi-configuration time-dependent Hartree (MCTDH) method~\cite{review_Beck2000_MCTDH,book_Meyer2009_MCTDH}, the variational multi-configurational Gaussian (vMCG) method~\cite{review_Richings2015_vMCG} and their extended versions~\cite{Valdes2018_C60_H2O,book_gonzalez2020_QD}. These methods have been shown to be highly efficient, although MCTDH normally requires a sum-of-product expression for the potentials to cover specific problems.

The Smolyak method~\cite{Smolyak1963_initial,Gradinaru2008_smolyak_math,Lasser2020_smolyak_math} has been used as an accurate and efficient tensor compression method for high-dimensional grids and applied successfully to obtain the vibrational spectra of molecules up to 12 degrees of freedom (DoF)~\cite{Avila2009_smolyak_initial,Avila2011_smolyak,Avila2015_smolyak,David2010_ElVibRot_Tnum,David2014_smolyak,david2018_clathrate,David2019_smolyak,Avila2019a_smolyak,Avila2019b_smolyak,Avila2020_smolyak}. At present, the approach is capable of treating a few tens of DoF, opening up a new direction for the improvement of quantum dynamics simulation~\cite{Avila2009_smolyak_initial,David2010_ElVibRot_Tnum,David2014_smolyak,Avila2015_smolyak,David2019_smolyak,Avila2019a_smolyak}. This method provides a way of selecting simultaneously basis functions and an associated sparse grid. So the number of basis functions and grid points will not scale exponentially with the system dimension. The Smolyak algorithm can be naturally embedded in a system-bath model (see \eq{eq_dir_expension}-\eq{eq_smolyak} below) to simulate a molecule or part of a molecule surrounded by a complex environment. The approach has no limitation on the formulation of the Hamiltonian and does not require the reconstruction of bath mode and system-bath coupling. However, we show here that converging the transition energies between system levels without converging the bath levels is rather difficult in the standard Smolyak scheme.

In this work, we propose a new Smolyak algorithm adapted to a system-bath separation (SSBS) to overcome this problem (see \fig{fig_clathrate}). To illustrate this method, we study a single H$_2$ molecule caged in a clathrate hydrate. A hydrogen clathrate hydrate is a typical host-guest system and is considered a promising hydrogen-storage candidate for future clean energy~\cite{Schuth_clathrate_H2_review,Mao2002_clathrate_H2_initial,Lee2005_clathrate_H2_THF}. The dynamics of the caged H$_2$ molecule is sensitive to the environment and highly quantum mechanical. 
A benchmark experiment~\cite{Ulivi2007_clathrate_exp} has confirmed the triplet splittings of both translational fundamental and rotational ($j$=1) transitions (T,R) of H$_2$ induced by the cage anisotropy~\cite{Bacic2006_clathrate_QD}.
However, investigating this system with full cage dynamics is challenging. We use SSBS to enable the simulation of the TR transitions for more than a hundred DoF. Our approach is tested using two different sets of cage normal modes (NM) to validate the method.

\section{Method}
In conventional quantum simulations, the wave function is expanded in primitive basis sets $B^i\equiv\{|b^{i}_{1}\rangle,|b^{i}_{2}\rangle,\cdots,|b^{i}_{N^{b^i}}\rangle \}$ ($i$$\in$$[1,n]$) as a full direct-product:
\begin{eqnarray}\label{eq_dir_expension}
|\Psi\rangle=\sum_{k_1=1}^{N^{b^1}}\cdots\sum_{k_n=1}^{N^{b^n}}C^{\bm{B}}_{\bm{k}}|b^{1}_{k_1}\rangle|b^{2}_{k_2}\rangle\cdots|b^{n}_{k_n}\rangle,
\end{eqnarray}
where $\bm{k}$$\equiv$$\{k_1,k_2,\cdots,k_n\}$, $\bm{B}$$\equiv$$\{B^1,B^2,\cdots,B^n\}$ and $C^{\bm{B}}_{\bm{k}}$ is the coefficient of the wave function expansion associated with the basis function, $|b^{1}_{k_1}\rangle|b^{2}_{k_2}\rangle\cdots|b^{n}_{k_n}\rangle$. $N^{b^i}$ is the size of the $B^i$ basis set. 

In the Smolyak method, the wave function is expressed instead as:
\begin{eqnarray}\label{eq_wf_smolyak}
|\Psi\rangle&=&\sum_{\mathscr{R}(\Bell)}D_{\Bell}|\psi^{\Bell}\rangle.
\end{eqnarray}
The Smolyak term, $|\psi^{\Bell}\rangle$, is projected on restricted basis sets  $B^i_{\ell}\equiv \{|b^{i}_{1}\rangle,|b^{i}_{2}\rangle,\cdots,|b^{i}_{N^{b^i}_{\ell_i}}\rangle\}$,
\begin{eqnarray}\label{}
|\psi^{\Bell}\rangle=\sum_{k_1=1}^{N^{b^1}_{\ell_1}}...\sum_{k_n=1}^{N^{b^n}_{\ell_n}}C^{\bm{B_\ell}}_{\bm{k}}|b^{1}_{k_1}\rangle |b^{2}_{k_2}\rangle\cdots|b^{n}_{k_n}\rangle,
\end{eqnarray}
as a small direct-product with respect to the usual full one (\eq{eq_dir_expension}), where $\bm{B_\ell}\equiv\{B^1_{\ell_1},B^2_{\ell_2},\allowbreak \cdots,B^n_{\ell_n}\}$, $\Bell\equiv\{\ell_1,\ell_2,\cdots,\ell_n\}$. $N^{b_i}_{\ell_i}$ is the size of the restricted basis $B^i_{\ell}$, defined as an arbitrarily increasing integer sequence function of $\ell_i$~\cite{Avila2011_smolyak,David2018_smolyak_clathrate,david2018_clathrate}. Usually, it is chosen as $A_i$+$B_i\ell_i$ ($A_i$$\geq$0, $B_i$$\geq$1, $A_i,B_i$$\in$$\mathbb{Z}$). The sum in \eq{eq_wf_smolyak} is restricted to specific Smolyak terms through the constraint, $\mathscr{R}(\Bell)$: $L$-$n$+1$\leq$$|\Bell|$$\leq L$, where $|\Bell|$=$\sum^n_{i=1} \ell_i$. The coefficient $D_{\Bell}$ is defined as $(-1)^{L-|\Bell|}(^{n-1}_{L-|\Bell|})$, where $(^{n-1}_{L-|\Bell|})$ is a binomial coefficient. The Smolyak parameter, $L$ (or $L^B$), is a constant that controls the approximation level for the $n$D-basis set and therefore its size. The total number of basis functions required is now reduced to $N^s_{L}$=$\sum_{\mathscr{R}(\Bell)}(\prod_{i=1}^{n}N^{b^i}_{\ell_i})$.
In terms of the tensor products in the Smolyak scheme~\cite{Smolyak1963_initial,David2014_smolyak}, the full $n$D-space can be represented as:
\begin{eqnarray}\label{eq_smolyak}
S_{L}=\sum_{\mathscr{R}(\Bell)}D_{\Bell} \cdot S^1_{\ell_1}\otimes S^2_{\ell_2}\otimes \cdots\otimes S^n_{\ell_n},
\end{eqnarray}
where $S^i_{\ell_i}$ ($i\in[1,n]$) can be the restricted basis sets $B^i_{\ell}$ or grids $\{Q^i_1,Q^i_2,\cdots,Q^i_{N^{q^i}_{\ell_i}}$\} and $N^{q_i}_{\ell_i}$ (the number of gaussian quadrature grid points) is defined similarly to $N^{b_i}_{\ell_i}$. This feature enables ones to define a unique Smolyak scheme simultaneously for the basis set and the grid, giving a sparse grid. For the grid, one can use another Smolyak parameter, $L^G$, usually larger than $L_B$. In this work, we always use $L^G=L^B$+1.


Using the standard Smolyak algorithm to tackle a quantum system, $s$, interacting weakly with a quantum environment or bath, $b$, leads to poor convergence of the energy-level transitions. Indeed, in the standard approach, the \emph{same} Smolyak parameter, $L$, is used for both system and bath parts. This is shown in the blue curves ($\bullet$) of \fig{fig_convergence} on a test with five H$_2$ modes and a water clathrate cage described using three normal modes. 
This originates from the restriction $\mathscr{R}(\Bell)$, leading to some basis functions describing the bath modes (clathrate cage) being present in the ground state expansion, but not in the excited states of the system (H$_2$), see supplement for a typical example~\cite{supplement}.
This imbalanced description of the ground state and the excited state levels leads to a lack of couplings in these excited states. This worsens as the bath is described using a growing number of DoF, leading to a large deviation in the energy transition. Increasing the value of $L$ can gradually reduce the importance of the missing coupling basis functions, which is the origin of the slow convergence of the transitions as a function of $L$.

\begin{figure} [ht!]
\begin{center}
\centering
\includegraphics[width=0.600\textwidth]{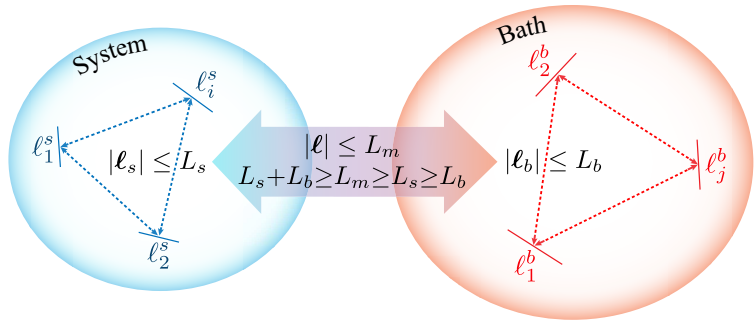}
\caption[]{(color online) An illustration of SSBS. The arrows denote the couplings between different modes (the solid lines) in the system and bath. The Smolyak terms in the system and bath are restricted by $|\Bell_s|\leq L_s$ and $|\Bell_b|\leq L_b$, respectively, while they are restricted overall by $L_m$ to ensure the system-bath coupling. See the text for more details.
}\label{fig_clathrate}
\end{center}
\end{figure}

The problem can also be attenuated by introducing a system-bath separation scheme, given the relatively weak coupling between the system and bath, affording an accurate description of the system part and a mild description of the bath part. Specifically, different Smolyak parameters, $L_s$ and $L_b$, are assigned to the system and bath parts, respectively. However, the convergence issue persists.
To solve the problem and ensure the full coupling of the system and bath (full-coupling SSBS), we introduce here an extra parameter $L_m$ covering the entire system with $L_m$=$L_s$+$L_b$, i.e.\ a direct-product between the system basis functions and the bath ones. This modification leads to a fast convergence of the energy-level transitions as shown in the black lines ($\times$) in   
\fig{fig_convergence}, which, however, requires a relatively large number of basis functions.
In order to better converge the lower states, we can further relax the condition on $L_m$ ($L_m$=$L_s$+$L_b$) to $L_s$$\leq$$L_m$$\leq$$L_s$+$L_b$, which finalizes the constraints (equivalent to $\mathscr{R}(\Bell)$) for general SBSS as
\begin{eqnarray}\label{eq_ssbs}
\mathscr{R}^{SSBS}(\Bell):\left\{
\begin{array}{l}
|\Bell_s| \leq L_s \\
|\Bell_b| \leq L_b \\
|\Bell| \leq L_m \\
 L_s+L_b\geq L_m\geq L_s\geq L_b,
\end{array}
\right.
\end{eqnarray}
where $|\Bell_s|$=$\sum_{i=1}^{n^s}\ell^s_i$ and $|\Bell_b|$=$\sum_{i=1}^{n^b}\ell^b_i$. It is worth noting that $|\Bell|$=$|\Bell_s|$+$|\Bell_b|$.
The coefficient $D_{\Bell}$ in \eq{eq_smolyak} is now obtained numerically by a direct expansion from the original Smolyak method~\cite{supplement,Smolyak1963_initial}.
The accuracy in describing the bath and system-bath coupling can be simply controlled by adjusting $L_b$ and $L_m$, respectively.
SSBS is equivalent to the original Smolyak scheme when $L_m$=$L_s$=$L_b$. An illustration of SSBS is shown in \fig{fig_clathrate}.

SSBS improves the convergence of the system part independently from the convergence of the bath part. Introducing $L_m$ guarantees the presence of ``necessary" basis functions for different states and so ensures a fast convergence of the energy-level transitions without a full convergence of the states. Thus it enables usage of a smaller basis set with respect to the original Smolyak scheme, as a smaller $L$ can be used. Various combinations of $L_m$, $L_s$, and $L_b$ can be chosen, according to the considered molecules, in which the full-coupling SSBS is one of the extreme cases (see the supplement~\cite{supplement}). A moderate combination, balancing the system-bath coupling and efficiency, is $C^0_{L_m}$: $L_s$=$L_m$-1, $L_b$=$\lceil L_m/2 \rceil$ ($\lceil~\rceil$ denotes the ceiling function).

\section{Results and discussion}
As an application of SSBS, we consider a hydrogen molecule in a $5^{12}$ sII clathrate hydrate cage~\cite{Mao2002_clathrate_H2_initial,Ulivi2007_clathrate_exp}. 
Numerous studies have contributed to the related quantum simulation~\cite{Colognesi2013_clathrate,Celli2013_clathrate,Powers2016_clathrate,David2018_smolyak_clathrate,David2019_smolyak,PNRoy2022_clathrate}, usually considering the cage to be rigid, namely, a quantum particle trapped within a confining potential.
According to the radial distribution of the oxygen atom from the cage center, the first three shells of sII cage contain 20, 20, and 36 water molecules, respectively~\cite{Powers2016_clathrate,david2018_clathrate}. 
In this work, only the first two shells are taken into account since the effects of the third shell are relatively weak~\cite{david2018_clathrate,David2018_smolyak_clathrate}. The water molecules in the second layer are assumed to be rigid to constrain the first layer in a solid phase environment. The dynamics of the water molecules in the inner layer is taken into account and described using fully delocalized rectilinear normal modes (first set). Thus the full dimension of the cage is 180 (9 per water molecule), where the 60 translational and 60 rotational DoF are crucial. As usual, the normal modes are sorted in terms of increasing frequencies, associated with the translation, rotation, bending, and stretching motions. The H$_2$ molecule is described by three translational DoF and two spherical angles. Furthermore, it is considered rigid, with a fixed inter-nuclear bond length~\cite{David2019_smolyak} at the equilibrium geometry of H$_2$ in the cage. The full Hamiltonian (H$_2$ and the first water layer) is given in the supplement~\cite{supplement}.
In terms of basis set, one-dimensional harmonic oscillator basis sets (HO) with $N^{b_i}_{\ell_i}$=1+$B_i\ell_i$ are associated with the NM of the cage and also to the 3 translational coordinates of H$_2$. Then, to describe H$_2$ rotations, we use a spherical harmonics basis ($Y^m_j$, with $j_{max}$=$B_i\ell_i$, $N^{b_i}_{\ell_i}$=$(1+B_i\ell_i)^2$) that can describe both ortho- and para- H$_2$~\cite{David2018_smolyak_clathrate,david2018_clathrate}.
For the grid points, we use Gauss-Hermite quadrature for the HO basis set and Lebedev grid points for the spherical spherical harmonics basis.
The potential energy surface (PES) considered is constructed using  
the flexible water-water SPC/Fw potential~\cite{Wu2006_potential_H2O} (one body and pair contributions) and the water-hydrogen SPC/E potential~\cite{Alavi2005_potential_H2_H2O}. The initial locations and orientations of the H$_2$O molecules are optimized by minimizing the H$_2$-cage interaction energy. The Cartesian atomic positions of the reference cage and the analytical expression of the full potential are given in the supplement~\cite{supplement}.



We present in Table \ref{table_num_basis} the number of required basis functions in SSBS ($N^{SSBS}_{L_m}$) when considering different DoF of the cage for the $C^0_{4}$ scheme: $L_m$=4, $L_s$=3, $L_b$=2 (note here $L^G_m$=5) and $B_s$=$B_b$=2. For H$_2$, the $B_i$ of the spherical harmonics basis is set equal to that of HO for convenience (they can be different generally). The results are compared with those of full-coupling SSBS, direct-product scheme $N^{dp}_{L_s,L_b}$=$\prod^{n_s}_{i=1}(N^{b_i}_{L_s})\prod^{n_b}_{i=1}(N^{b_i}_{L_b})$ and standard Smolyak algorithm $N^{s}_{L_m}$. 
Since the same basis function can be present in several Smolyak terms, $N^{SSBS}_{L_m}$ and $N^{s}_{L_m}$ can technically be made even smaller by removing the duplication~\cite{supplement}. We should note, for the Smolyak method, that the $L_m$ chosen here is actually far too small to obtain full convergence. 
\begin{table} [ht!]
\begin{center}
\begin{tabular}[t]{c|ccccccl}
\specialrule{0.10em}{0em}{0.15em}
$C^0_{4}$ & $N^{dp}_{L_s,L_b}$ & $N^{s}_{L_m}$ & $N^{SSBS}_{L_m}$ & $N^{SSBS}_{\text{full coupling}}$ \\
\specialrule{0.05em}{0em}{0em}
($L_m$,$L_s$,$L_b$) & / & (4,4,4) & (4,3,2) & (5,3,2) \\
\begin{tabular}{@{}c@{}}5+3D \end{tabular} & 2100875 & 122740 & 104580 & 347034 \\ 
\begin{tabular}{@{}c@{}}5+60D \end{tabular} & 1.5$\times10^{46}$ & 2624892199 & 179516184 & 1003272294 \\ 
\begin{tabular}{@{}c@{}} 5+120D \end{tabular} & 1.2$\times10^{88}$ & 65600962720 & 1365782634 & 7732041654 \\
\specialrule{0.10em}{0em}{0em}
\end{tabular}
\caption{Number of basis functions required when considering caged H$_2$ with different DoF of the cage in conventional direct-product scheme ($N^{dp}_{L_s,L_b}$), standard Smolyak algorithm ($N^{s}$), SSBS with scheme $C^0_{L_m}$ ($N^{SSBS}_{L_m}$) at $L_m$=4, $L^G_m$=5 and the related full-coupling SSBS. 
}\label{table_num_basis}
\end{center}
\end{table}


\begin{figure} [ht!]
\begin{center}
\centering
\includegraphics[width=0.750\textwidth]{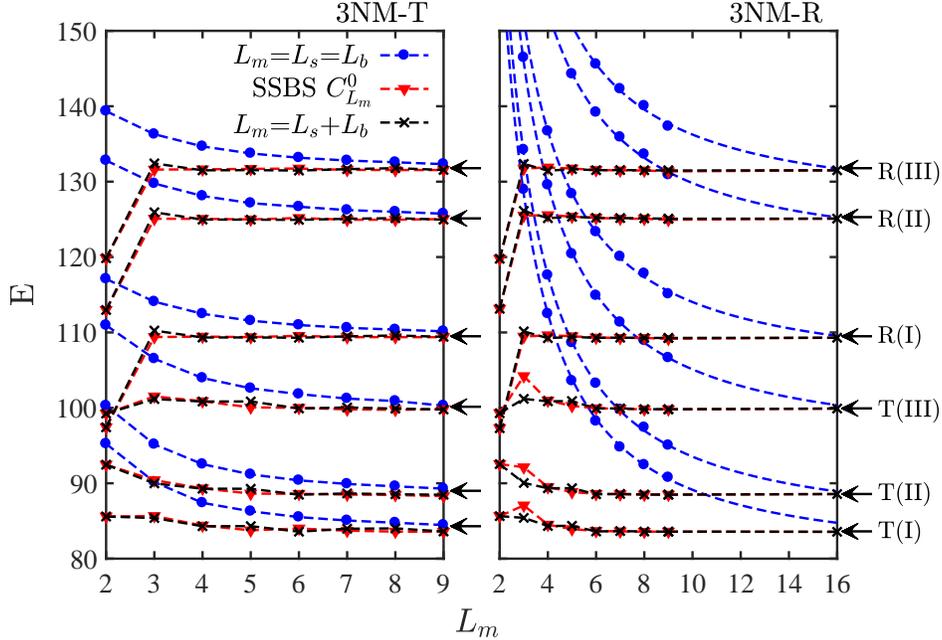}
\caption[]{(color online) The convergence of TR states of H$_2$ as a function of $L_m$ in standard Smolyak algorithm (blue $\bullet$) and SSBS $C^0_{L_m}$ at $B_s$=$B_b$=2 (red $\blacktriangledown$) as indicated. The results of full-coupling SSBS (black $\times$) are also shown for a comparison. The benchmark results of direct-product scheme are marked with black arrows (\contour{black}{$\leftarrow$}). The dashed lines are guide for the eyes. The test is performed for 3 NM involving mainly translational (3NM-T) or rotational (3NM-R) DoF of H$_2$O molecules to ensure the convergence in general.}
\label{fig_convergence}
\end{center}
\end{figure}

The SSBS simulation is performed with our program \textsc{ElVibRot}~\cite{David2010_ElVibRot_Tnum,David_ElVibRot_TnumTana,ElVibRot_MPI}, a Smolyak method quantum simulation package working with curvilinear coordinates. The diagonalization is performed with a block-Davidson procedure. It has been verified in a series of quantum simulations~\cite{David2010_ElVibRot_Tnum,David2014_smolyak,david2018_clathrate,David2018_smolyak_clathrate,David2019_smolyak}.
As an application of our SSBS approach, we present in \fig{fig_convergence} convergence tests for an H$_2$ clathrate hydrate, considering 3 NM of the cage associated mainly with the translational (3NM--T) or rotational (3NM--R) water DoF. 
The tests are performed at $B_s$=$B_b$=2 as a function of $L_m$ for different constraints $\mathscr{R}^{SSBS}(\Bell)$ to cover the behaviour of different combinations. The results shown in \fig{fig_convergence} for $C^0_{L_m}$ and full-coupling SSBS are also compared with those of a standard Smolyak algorithm and a direct-product scheme. More details can be found in the supplement~\cite{supplement}. 
 
With the standard Smolyak algorithm (blue dashed line in \fig{fig_convergence}), the transition energy levels show a very slow convergence as a function of $L_m$.
This convergence issue is even worse when the rotational NM of water (3NM--R) are considered. In this case, the standard Smolyak results have to be extrapolated to convergence.
When aiming at converged TR transitions of H$_2$, a large portion of basis functions is not only ``wasted" for a rigorous description of the cage part, but also causes the deviation of the energies of TR states as mentioned above. Therefore, at least $L_m$=9 is required to converge the TR states of H$_2$ with a standard Smolyak method for the 3NM--T model, leading to $N^s_{9}$=74387784 basis functions.
SSBS overcomes this problem for the TR transitions, exhibiting very fast convergence with $L_m$. All the tested combinations show similar robust convergence.
This suggests that the actual $L_m$ required for a fair tensor compression with Smolyak method for TR transitions is smaller than expected. Indeed, $L_m$=4 seems sufficient (see Table \ref{table_num_basis}).
These convergence properties hold when more NM are taken into account. Therefore, it is possible to achieve simulations of much larger systems.

\begin{figure} [ht!]
\begin{center}
\includegraphics[width=0.90\textwidth]{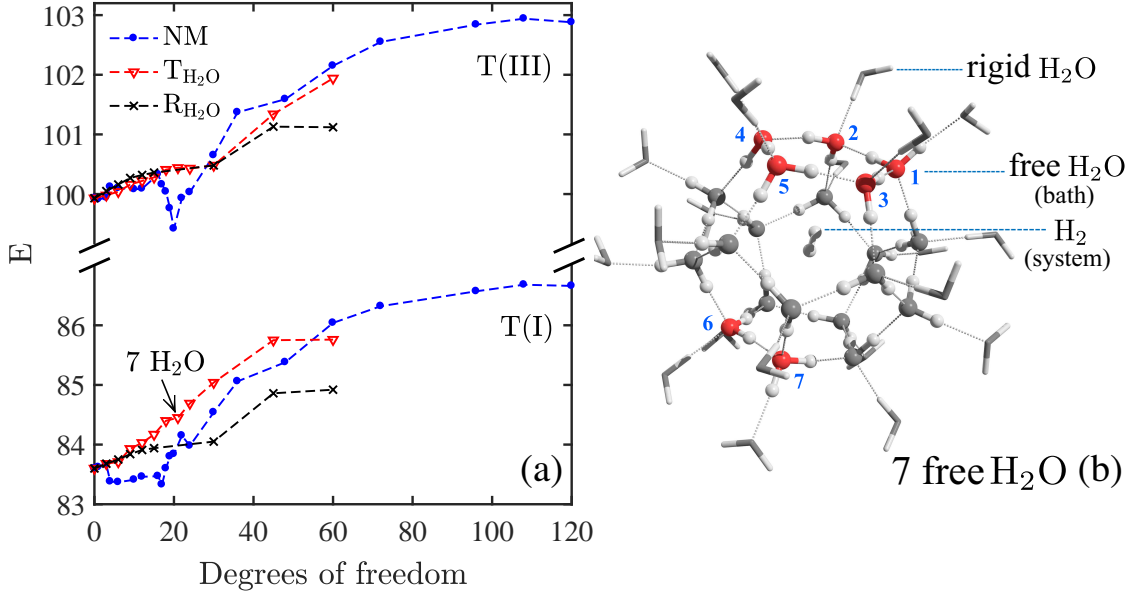}
\caption{(color online) Panel (a) shows the T(I) and T(III) energies as a function of number of DoF for the delocalized NM (first set),  T$_{\text{H}_2\text{O}}$, or R$_{\text{H}_2\text{O}}$ modes (second set). Panel (b) shows the initial locations of the first 7 H$_2$O molecules chosen (colored) for T$_{\text{H}_2\text{O}}$ and R$_{\text{H}_2\text{O}}$ as indicated in panel (a). 
} \label{fig_TR_H2O}
\end{center}
\end{figure}

By increasing the number of NM in the simulation with SSBS ($C^0_{L_m}$) until convergence (around 100 NM, see the blue dashed curves in \fig{fig_TR_H2O}(a)), we obtain the TR transitions of H$_2$, denoted as T(I, II, III) and R(I, II, III), respectively. The results for rigid cage, with 60 and 120 NM obtained at $C^0_4$ are shown in Table \ref{table_TR_nm} (named 60NM and 120NM, respectively). The convergence of each simulation is confirmed by increasing the $L_m$ in $C^0_{L_m}$. The splitting of TR levels, $\Delta$T=T(III)-T(I) and $\Delta$R=R(III)-R(I), are also shown as a reference. The simulation time for 120 NM is around 7.8 hours in parallel computation on 15 nodes with 80 cores per node. The results are compared with the experimental data and typical previous simulation results~\cite{Bacic2008_clathrate,david2018_clathrate}, in which the Ref~\cite{Bacic2008_clathrate} used the same SPC/E potential as this work but different cage configurations. The results obtained in Ref~\cite{david2018_clathrate} accounted for the quantum effects of the overall rotation of the water molecules with an adiabatic-QDMC approach and they show a weak effect on the transition energies with respect to a rigid cage. Overall, our TR energies obtained agree well with the experiment, reproducing the triplet splittings of TR states.
The T(III) and R(I) states have very good agreement with the experiment. However, $\Delta$T and $\Delta$R are under- and over-estimated, respectively. The deviation is likely due to the PES used in our simulation~\cite{Zahra2015_3bodyH20_po,VALIRON08}.

Furthermore, we observe in Table 2 that the published calculation using a rigid cage~\cite{Bacic2008_clathrate} is in better agreement with the experimental data than our current calculations, even with a rigid cage (first line of Table 2), although both calculations use the same water-hydrogen SPC/E potential~\cite{Alavi2005_potential_H2_H2O}. This effect is due to the shape of the water cage which is different between the two studies. Indeed, to compute the normal modes, the water positions of the first shell have been optimized. More generally, we can argue that such difference is coming from the full potential (water-hydrogen and water-water contributions) which is not perfect. Our current approach is able to consider explicitly the large number of degrees of freedom and therefore to probe directly the quality of a potential.
\begin{table} [ht!]
\begin{center}
\begin{tabular}[t]{ccccccccc}
\specialrule{0.10em}{0em}{0em}
 & T(I) & T(II) & T(III) & R(I) & R(II) & R(III) & $\Delta$T & $\Delta$R\\
\specialrule{0.05em}{0em}{0em}
rigid & 83.6 & 88.5 & 99.8	 &	109.3 & 125.0 & 131.5 & 16.3 & 22.3 \\
{\bf 60NM } & {\bf 86.1} & {\bf 90.9 } & {\bf 102.3}	&	{\bf 110.1} & {\bf 125.5} & {\bf 132.1} & {\bf 16.1 } & {\bf 22.0 } \\
{\bf 120NM} & {\bf 86.7} & {\bf 91.6 } & {\bf 102.9}	&	{\bf 110.8} & {\bf 125.0} & {\bf 132.0} & {\bf 16.2 } & {\bf 21.2 } \\
60T$_{\text{H}_2\text{O}}$ & 85.9 & 90.4 & 102.1	&	110.3 & 125.3 & 132.1 & 16.2 & 21.8 \\
60R$_{\text{H}_2\text{O}}$ & 84.9 & 89.7 & 101.1	&	111.3 & 125.4 & 132.4 & 16.2 & 21.1 \\
{\bf \footnotesize 120(T+R)$_{\text{H}_2\text{O}}$} & {\bf 86.4} & {\bf 91.3} & {\bf 102.8}	&	{\bf 111.2} & {\bf 125.3} & {\bf 132.4} & {\bf 16.3} & {\bf 21.2} \\
\specialrule{0.05em}{0em}{0em}
exp. & 71.0 & 80.2 & 101.1 &	110.0 & 116.5 & 122.1 & 30.1 & 12.1 \\ 
Ref.\cite{Bacic2008_clathrate} & 74.5 & 74.7 & 97.5 & 109.0 & 119.4 & 128.1 & 23.1 & 19.2\\
Ref.\cite{david2018_clathrate} & 80.5 & 85.1 & 107.7 & 108.6 & 117.8 & 124.1 & 27.2 & 15.5 \\
\specialrule{0.10em}{0em}{0em}
\end{tabular}
\caption{The energies (in cm$^{-1}$) of TR transitions, obtained with SSBS at $C^0_4$, with rigid cage, 60 or 120 delocalized normal modes (named 60NM and 120NM). The results of the calculation considering localized normal modes, 60T$_{\text{H}_2\text{O}}$, 60R$_{\text{H}_2\text{O}}$ and full 120 (T+R)$_{\text{H}_2\text{O}}$ dynamics are also shown. They are compared with those of experiment and typical previous works~\cite{Bacic2008_clathrate,david2018_clathrate}.}\label{table_TR_nm}
\end{center}
\end{table}

Specifically, the fair reproduction of TR transitions with the rigid cage model suggests that we have obtained suitable initial H$_2$O locations through our geometry optimization. With 60 and 120 NM, the $\Delta$T, $\Delta$R, and R levels are not observably influenced by H$_2$O dynamics, as was the case for other simulations~\cite{david2018_clathrate}, while the T levels are overall lifted by roughly 3 cm$^{-1}$.
In \fig{fig_TR_H2O}(a) we show the T(I) and T(III) transitions as a function of NM ranging from 0 (rigid cage) to 120 DoF. The increase of T energies slows down around 70 NM, reaching convergence already around 100 NM. 
When examining the full energy levels of H$_2$ clathrate hydrate, we found that the coupling of the excitation of H$_2$ with that of NM forms a series of states around the TR levels. For instance, the coupling of T(I) with the excitation of NM forms several levels between R(I) and R(II). Besides, the convergence of R levels only requires just a few basis functions. It indicates an almost free rotation of H$_2$ in the cage.

Instead of the delocalized normal modes employed above, we could use another set (the second set) of normal modes localized around water molecules. Those modes are sorted in terms of increasing frequencies for each water molecule, but, one has some freedom to select the water molecule ordering. Therefore, the first few H$_2$O are chosen in the same cage pentagon as shown in \fig{fig_TR_H2O}(b) (see the supplement~\cite{supplement} for the full selection). They are named, respectively, T$_{\text{H}_2\text{O}}$,  R$_{\text{H}_2\text{O}}$ and (T+R)$_{\text{H}_2\text{O}}$ when rotation or translation or both motions is taken into account (see Table \ref{table_TR_nm} and \fig{fig_TR_H2O}). 
The TR transition energies at 60T$_{\text{H}_2\text{O}}$, 60R$_{\text{H}_2\text{O}}$ and full 120(T+R)$_{\text{H}_2\text{O}}$ obtained at $C^0_{4}$ are shown in Table \ref{table_TR_nm}. The results of 60T$_{\text{H}_2\text{O}}$ and 120(T+R)$_{\text{H}_2\text{O}}$ are very close to those of 60NM and 120NM as expected, since the first 120 NM mainly consist of 60 translational and 60 rotational DoF. It suggests that couplings in the cage are correctly taken into account. This further confirms the convergence of our simulation.

\section{Conclusion}
In conclusion, we have proposed an SSBS method to address the TR transitions of H$_2$ molecule in an sII clathrate hydrate cage considering the cage dynamics. In terms of system-bath separation, this Smolyak algorithm method overcomes the issues with eigenstate convergence and provides rapid convergence for the transitions of the ``system" part without converging the ``bath" part transitions. It significantly pushes the limit of the Smolyak algorithm to over a hundred DoF for some molecules. 
The obtained results agree well with experimental data.
The simulations performed in this work use small-scale computing capabilities, implying that SSBS can be extended to even more DoF when deployed on large-scale architectures. 
By using $L_m$$>$$L_s$=$L_b$ in \eq{eq_ssbs}, this method is generally helpful for dealing with similar convergence issues when using the Smolyak algorithm in quantum simulations.

In terms of applications, the proposed Smolyak scheme, SBBS, is well-adapted to study a molecule coupled to an environment. The present work shows this for weak couplings but we plan to extend to system-bath systems that show stronger couplings. Atoms or molecules in fullerenes~\cite{Felker2016_C60_H2O,Valdes2018_C60_H2O,Carrillo2021_C60_H2O,Xu2022_C60_H2O}, carbon nanotubes~\cite{Manel2021_H2_tube,Review_Castells2021_Nanotubes_Confinement}, clathrate (aluminosilicates)~\cite{Zhong2019_noble_gas_clathrate}, metal-organic frameworks~\cite{FitzGerald2008_H2_MOF5} are all accessible with this new approach.
For more complex systems involving several molecules encapsulated in a nanoscale environment, the calculations are formally doable. However, the size of the basis set might be very large and possibly require some adjustments or coupling of the Smolyak scheme for the bath to other approaches for the system, like the one used to study two H$_2$ molecules in hydrate clathrate~\cite{Felker2019_H2_clathrate}.

\begin{acknowledgement}
A.C. acknowledges the funding support from E-CAM European Centre of Excellence, European Union's Horizon 2020 research and innovation program under Grant No. 676531.
We acknowledges the computational resource of the {\it styx} in
Institut de Chimie Physique, Universit\'{e} Paris-Saclay, the {\it mandelbrot} in Maison de la Simulation, CEA-Saclay, and the {\it JUWELS} in J\"{u}lich Supercomputing Centre provided by Dr.\ Alan O'Cais in E-CAM. 
\end{acknowledgement}

\begin{suppinfo}
\begin{itemize}
  \item Supplement\_SSBS.pdf: additional details on the Smolyak algorithm, SSBS, and the setup and test of the simulation.
\end{itemize}
This information is available free of charge via the Internet at \href{http://pubs.acs.org}{http://pubs.acs.org}
\end{suppinfo}

\bibliography{achemso-demo}

\end{document}